# Black Holes, Entropy and the Third Law

A. J. Meyer, II
*International Scientific Projects, Inc.*
*PO Box 3477*
*Westport Connecticut, 06880 USA*
*Email:* ajmeyer@InternationalScientificProjects.org

**Abstract**
There would be a perfect correspondence between the laws of classical thermodynamics and black hole thermodynamics, except for the apparent failure of black hole thermodynamics to correspond to the Third Law. The classical Third Law of Thermodynamics entails that as the absolute temperature, T, approaches zero, the entropy, S, also approaches zero.

This discussion is based upon part of the work published by the author in 1995 that demonstrated that the most general form of the classical Third Law of Thermodynamics is satisfied by treating the area of the inner-event horizon as a measure of negative-entropy (negentropy).

**Keywords:** entropy, negentropy, Third Law, Zeroth Law, First Law, Second Law, thermodynamics, Planck, primaton, Kerr, Schwarzschild, black hole, Super Spin Model (SSM), inner event horizon, temperature, surface gravity.

## Background

There would be a perfect correspondence between the laws of classical thermodynamics and black hole thermodynamics, except for the apparent failure of black hole thermodynamics to correspond to the classical Third Law. The classical Third Law of Thermodynamics entails that as the absolute temperature, T, approaches zero, the entropy, S, also approaches zero.

This paper is based upon a portion of the work, ref. [11], published by the author in 1995 that demonstrated that the most general form of the classical Third Law of Thermodynamics is satisfied by treating the area of the inner-event horizon as a measure of negative-entropy (negentropy) and adding it to (in effect, subtracting it from) the outer-event horizon area from which the standard Bekenstein-Hawking (BH) entropy is calculated.

## Kerr family of Black Holes

The Kerr uncharged family [(1)] is determined by just two parameters: M and **J**, where M is the total mass of the black hole and **J** is its net angular momentum. The Kerr family, [1], is represented by the following metric:

$$ds^2 = (2R_g r - \sigma^2)du^2/\sigma^2 - 2aR_z^2 \sin^2\theta d\phi du/\sigma^2 + R_\phi^4 \sin^2\theta d\phi^2/\sigma^2 \qquad (1)$$
$$+ 2dudr - 2a\sin^2\theta dr d\phi + \sigma^2 d\theta^2.$$

Where:

$$Mc^2 \equiv \text{Total energy of black hole,} \qquad (2)$$
$$R_g \equiv GM/c^2 \equiv \text{Gravitational radius of black hole,} \qquad (3)$$
$$J \equiv |\mathbf{J}| \equiv aMc \equiv \text{Scalar value of black hole's angular momentum or}$$
$$a \equiv \text{Black hole's specific angular momentum radius} \equiv J/Mc, \qquad (4)$$
$$\sigma^2 \equiv r^2 + a^2\cos^2\theta, \qquad (5)$$
$$R_z^2 \equiv 2R_g r, \qquad (6)$$
$$\gamma^2 \equiv R_z^2 - r^2 - a^2, \qquad (7)$$
$$R_\phi^4 \equiv (r^2 + a^2)^2 + \gamma^2 a^2 \sin^2\theta, \qquad (8)$$

such that:

$$-\infty \le r \le +\infty, \qquad (9)$$
$$-\infty \le u \le +\infty,$$

[(1)] The analysis in this paper applies equally as well to the more general Kerr-Newman metric.

$$0 \le \phi \le 2\pi,$$
$$0 \le \theta \le \pi.$$

Note, there exist two non-negative *event horizon radii,* $r_+$ and $r_-$, which are the conjugate solutions of the equation: $\gamma^2(r) = 0$, which are:

$$r_\pm = R_g \pm [R_g^2 - a^2]^{1/2}. \quad (10)$$

The horizon radii can be re-written as:

$$r_\pm = R_g[1 + sin(\pm \Phi)] \Rightarrow \quad (11)$$

$$a^2(\Phi) = r_+r_- = R_g^2 cos^2\Phi \text{ and } r_+ + r_- = 2R_g, \quad (12)$$

with boundary conditions:

$$0 \le r_- \le R_g \le r_+ \le 2R_g, \quad (13)$$
$$-\pi/2 \le \Phi \le \pi/2.$$

The event horizons' *rotation-offset radii* are:

$$\sigma_\pm = R_g [(1 \pm sin\Phi)^2 + cos^2\Phi cos^2\theta]^{1/2}. \quad (14)$$

The *radii of gyration* of the event horizons are:

$$R_{z\pm}(\Phi) = [2R_g r_\pm]^{1/2} = [r_\pm^2 + a^2]^{1/2} = R_g[2(1 \pm sin\Phi)]^{1/2}. \quad (15)$$

The *dual event horizon rotation rates* are:

$$\Omega_\pm(\Phi) = cd\phi/du_\pm = -g_{u\phi}(\Phi)c/g_{\phi\phi}(\Phi) = ac/(R_{z\pm})^2 = c\, cos\Phi/[2R_g(1 \pm sin\Phi)]. \quad (16)$$

For a constant M, one can see that the *moments of inertia* are:

$$I_\pm(\Phi) = MR_{z\pm}^2, \quad (17)$$

since:

$$I_\pm\Omega_\pm = J(\pm\Phi) = |\mathbf{J}(\pm\Phi)| \equiv J_\pm = a(\pm\Phi)Mc = GM^2 cos\Phi/c = R_g Mc\, cos\Phi. \quad (18)$$

Where, the black hole's specific angular momentum radius, 'a' is parameterized in terms of M and $\pm\Phi$, so that:

$$a = a(\pm\Phi) = |\mathbf{J}(\pm\Phi)|/Mc = GM cos\Phi/c^2 = R_g\, cos\Phi. \quad (19)$$

The angle '$\Phi$' is therefore a parametric measure of the magnitude of the black hole's angular momentum. That is:

$$J(M, \pm \Phi) = |\mathbf{J}(M, \pm \Phi)| = GM^2 cos\Phi/c = a(\pm\Phi)Mc = R_g Mc\, cos\Phi. \quad (20)$$

As is easily seen, $\Phi = \pm 0$ implies a maximally rotating (extreme) Kerr black hole and $\Phi = \pm \pi/2$ implies a static Schwarzschild black hole.

**Event Horizon Areas '$A_\pm$'**

For the parametric range, $0 \le \Phi \le \pi/2$, the area '$A_+$' of the outer event horizon of a Kerr black hole is:

$$A_+ \equiv A(+\Phi) = \int_0^{\pi} \int_0^{2\pi} \sqrt{(g_{\phi\phi}g_{\theta\theta})}\, d\phi d\theta = 8\pi R_g[R_g + (R_g^2 - (J/Mc)^2)^{1/2}] =$$
$$8\pi R_g r_+ = 8\pi R_g^2 [1 + sin(+\Phi)] = 4\pi R_{z+}^2 \quad (21)$$
$$\Rightarrow$$
$$8\pi R_g^2 \le A_+ \le 16\pi R_g^2. \quad (22)$$

And for the parametric range, $-\pi/2 \le \Phi \le 0$, the area, '$A_-$' of the inner event horizon of a Kerr black hole is:

$$A_- \equiv A(-\Phi) = \int_0^{\pi} \int_0^{2\pi} \sqrt{(g_{\phi\phi}g_{\theta\theta})}\, d\phi d\theta = 8\pi R_g[R_g - (R_g^2 - (J/Mc)^2)^{1/2}] =$$
$$8\pi R_g r_- = 8\pi R_g^2 [1 + sin(-\Phi)] = 4\pi R_{z-}^2 \quad (23)$$
$$\Rightarrow$$
$$0 \le A_- \le 8\pi R_g^2. \quad (24)$$

Now form a "net" area 'A', where:

$$A \equiv A_+ - A_- = 4\pi(R_{z+}{}^2 - R_{z-}{}^2) = 16\pi R_g{}^2 sin\Phi. \quad (25)$$

Note, from eqs.(22), (24) & (25) that:

$$A \geq 0. \quad (26)$$

Hence, 'A' can be written as:

$$A = 16\pi R_g{}^2 sin|\Phi|. \quad (27)$$

**The Horizon Surface Gravities '$\kappa_\pm$'**

The two surface gravities '$\kappa_\pm$' over the two event horizons are represented by the two conjugate roots of:

$$\kappa_\pm = c^2(R_g{}^2 - (J/Mc)^2)^{1/2}/2R_g[R_g \pm (R_g{}^2 - (J/Mc)^2)^{1/2}] = c^2 sin(\pm\Phi)]/2R_g[1 + sin(\pm\Phi)] \quad (28)$$

$$\Rightarrow$$

$$-\infty \leq \kappa_- \leq 0 \leq \kappa_+ \leq c^2/2R_g. \quad (29)$$

By eq.(15), '$\kappa_\pm$' can be written in terms of the *radii of gyration* '$R_{z\pm}$' of the event horizons, as:

$$\kappa_\pm = GM sin(\pm\Phi)/R_{z\pm}{}^2 = c^2 R_g sin(\pm\Phi)/R_{z\pm}{}^2. \quad (30)$$

**The Horizon Temperatures '$T_\pm$'**

The horizon temperatures '$T_\pm$' over the two event horizons are calculated by:

$$T_\pm = \hbar\kappa_\pm/2\pi kc, \quad (31)$$

then by eqs.(30), (31), [(2)]

$$T_\pm = \hbar GM sin(\pm\Phi)/2\pi kcR_{z\pm}{}^2 = \hbar c\, sin(\pm\Phi)/4\pi kR_g[1 + sin(\pm\Phi)] = \hbar cR_g sin(\pm\Phi)/2\pi kR_{z\pm}{}^2. \quad (32)$$

where 'k' is Boltzman's constant, i.e. *the fundamental quantum of entropy.*

**Bekenstein-Hartle-Hawking Temperature '$T_{BH}$'**

The Bekenstein-Hartle-Hawking Temperature '$T_{BH}$' is generally calculated over the outer event horizon by using only the positive root of eq.(28), which gives:

$$T_{BH} = \hbar\kappa_+/2\pi kc = \hbar c\, sin|\Phi|/4\pi kR_g[1 + sin|\Phi|]. \quad (33)$$

So when $\Phi = \pi/2$, the Bekenstein-Hartle-Hawking temperature represents the temperature over the outer event horizon of a Schwarzschild black hole. That is:

$$T_{BH}(\pi/2) = \hbar\kappa_+/2\pi kc = \hbar c/8\pi R_g = \hbar c^3/8\pi kGM. \quad (34)$$

**Bekenstein-Hawking entropy '$S_{BH}$'**

The work done by Bekenstein and Hawking has established that:

$$S_{BH} \equiv S_+ = kc^3 A_+/4\hbar G = 2\pi kGM^2[1 + sin|\Phi|]/\hbar c. \quad (35)$$

See references: [3], [4], [5], [6], [7] and [16].

**Entropy and the Third Law**

The classical, or strong version of the Third Law of thermodynamics, states that as the temperature approaches zero, the entropy also approaches zero.

The work done by Smarr [2], Bardeen, Carter and Hawking [3], Bekenstein [4], [5] and Hawking [6], [7], has demonstrated that except for the Third Law, there is consistency between Black Hole Thermodynamics and classical Thermodynamics (see Wald [8]). There has also been some recent work addressing the apparent failure of Black Hole Thermodynamics to satisfy the Third Law, see R´acz [9] and Liberati, Rothman Sonego [10].

A basic hypothesis in the Super Spin Model (SSM), Meyer [11], [12], [13], is that: the most general form of the Third Law of Thermodynamics is satisfied by treating the area of the inner-event horizon as a measure of negative-entropy (negentropy), and adding it to (in effect, subtracting it from) the Bekenstein-Hawking entropy associated with the outer event horizon.

---

(2) Note: $T(\Phi)_{\lim(\Phi \to -\pi/2)} = -\infty$ °K

By rewriting the Bekenstein-Hawking (BH) entropy, eq.(35), in Planck units, we get:

$$S_{BH}(\Phi) \equiv S_+ = kc^3 A_+/4\hbar G = 2\pi k N_\varpi^2 [1+ sin|\Phi|], \quad (36)$$

Where the black hole's mass '$N_\varpi$' in Planck units is:

$$N_\varpi \equiv M/m_\varpi = R_g/r_\varpi, \quad (37)$$

where the Planck mass '$m_\varpi$' is:

$$m_\varpi = (\hbar c/G)^{1/2},$$

and the Planck radius '$r_\varpi$' is:

$$r_\varpi = (\hbar G/c^3)^{1/2}.$$

Note:

$$\hbar = r_\varpi m_\varpi c$$

Now, for maximally spinning black holes:

$$R_g = J/Mc \Rightarrow \Phi = 0 \Rightarrow A_+ = 8\pi R_g^2, \text{ by eq.(19) \& eq.(21).} \quad (38)$$

It then follows, from eqs.(36) & (37) that the Bekenstein-Hawking entropy, $S_{BH}(\Phi)$, for all ***extreme*** black holes much greater than the Planck mass,[(3)] is:

$$S_{BH}(0) = 2\pi kGM^2/\hbar c = 2\pi k N_\varpi^2 >> 0, \quad (39)$$

However, from eq.(28) & eq.(31), the Bekenstein-Hawking temperature for the outer event horizon of an extreme Kerr black hole [i.e.($\Phi$=0)] is:

$$T_{BH}(0) = \hbar\kappa_+(0)/2\pi kc = \hbar c^2 sin(0)/4\pi kcR_g[1 + sin(0)] = 0. \quad (40)$$

The above formulation for Bekenstein-Hawking entropy obviously fails to satisfy the strong version of the Third Law, since even at zero temperature, non-microscopic extreme black holes still have enormous entropy. That is, even at zero temperature, there exists a large non-zero outer event horizon area, such that its associated entropy is equal to one half its maximum possible entropy, $S(\pi/2) = 4\pi k N_\varpi^2$.

**The Strong Third Law is Satisfied**

**Premise:**
The Strong Third Law is satisfied if the net area 'A' of a Kerr black hole is calculated by:

$$A \equiv A_+ - A_- \quad (41)$$

The net entropy [(4)], by eqs. (25), (37) & (41) is then:

$$S = kc^3A/4\hbar G = 4\pi kGM^2 sin\Phi/\hbar c = 4\pi R_g^2 kc^3 sin\Phi/\hbar G = 4\pi k(R_g/r_\varpi)^2 sin\Phi, \quad (42)$$

or in Planck units,

$$S = 4\pi k N_\varpi^2 sin\Phi \Rightarrow \quad (43)$$

$$S(0) = 0, \quad (44)$$

$$S(\pi/2) = 4\pi k N_\varpi^2, \quad (45)$$

where:

$$\Phi = cos^{-1}(J/R_gMc), \quad (46)$$

then the Strong Third Law is satisfied. Since, for maximally rotating extreme black holes, when $\Phi_\pm = 0$, then

$$R_g = J/Mc \text{ and } T_\pm = \pm\hbar c(R_g^2 - (J/Mc)^2)^{1/2}/4\pi kR_g[R_g \pm (R_g^2 - (J/Mc)^2)^{1/2}] = 0. \quad (47)$$

---

(3) Of course, for primatons, i.e. extreme Planck mass ($m_\varpi$) Kerr black holes, the Bekenstein-Hawking entropy is:

$$S_{BH} = 2\pi k(M/m_\varpi)^2 = 2\pi k N_\varpi^2 = 2\pi k, \text{ since } M = m_\varpi.$$

See Meyer [11], [14]

(4) This means that the conventional Bekenstein-Hawking outer-event horizon entropy for a rotating black hole, spinning with an angular momentum of around 86.6% of the extreme maximum, will be about 50% greater than the net entropy of the same rotating black hole, as calculated by using both its event horizons. See Figure 1.

This extension appears to make the correspondence between Black Hole Thermodynamics and Classical Thermodynamics complete. Since, for maximally rotating (extreme) black holes,

$$J = R_g Mc \Rightarrow \Phi = 0 \Rightarrow S = 0, \text{ when } T_\pm = \hbar\kappa_\pm/2\pi kc = 0. \quad (48)$$

This result is completely consistent with the most stringent interpretation of the Third Law, which has been bent and stretched a bit, by trying to make it appear consistent with the rest of Black Hole Thermodynamics.

This modification of the amount of entropy produced by black holes may be of some importance in astrophysics, illustrated by the work done by Kephart and others on the mergers of black holes and the entropy budget of the Universe. See Figure 1 and reference [15].

**Consistency with the other Fundamental Laws of Thermodynamics**

By treating the net area of a black hole as the difference in the areas of the outer and inner event horizons, it appears that the Third Law of black hole thermodynamics is in complete agreement with the strong version of the Third Law in classical thermodynamics. However such a change may cause some inconsistency with the rest of thermodynamics. The following analysis shows no such conflict.

It can be shown that $\partial\mathbf{r}_+ \bullet \partial\mathbf{r}_- = 0$. Therefore, by using eq. (15), one can form a net square of the radius of gyration by:

$$R_z^2 = R_{z+}^2 + R_{z-}^2 = 4R_g^2. \quad (49)$$

By analogy to '$\kappa_\pm$' in eq.(30), a "net surface" gravity '$\kappa$' can be formed, so that:

$$\kappa = c^2 R_g \mathit{sin}\Phi/R_z^2 = c^2 \mathit{sin}\Phi/4R_g. \quad (50)$$

Then, the net temperature 'T' is reckoned as:

$$T = \hbar\kappa/2\pi kc = \hbar c\,\mathit{sin}\Phi/8\pi kR_g = \hbar c^3\,\mathit{sin}\Phi/8\pi kGM = m_\varpi c^2 \mathit{sin}\Phi/8\pi kN_\varpi. \quad (51)$$

Now, repeating eq.(43):

$$S = 4\pi kN_\varpi^2 \mathit{sin}\Phi. \quad (52)$$

Then:

$$TS = m_\varpi c^2 N_\varpi \mathit{sin}^2\Phi/2 = Mc^2 \sin^2\Phi/2, \quad (53)$$

since $M = N_\varpi m_\varpi$.

Now let:

$$M_\pm \equiv M(1\pm \mathit{sin}\Phi)/2. \quad (54)$$

Then:

$$M = M_+ + M_- . \quad (55)$$

Then from eqs.(15), (16) & (55): (5)

$$J_\pm = M_\pm\Omega_\pm R_{z\pm}^2 = R_g Mc\, \mathit{cos}\Phi(1\pm \mathit{sin}\Phi)/2 \quad (56)$$

$$\Rightarrow$$

$$J_+\Omega_+ = J_-\Omega_- = Mc^2 \mathit{cos}^2\Phi/4. \quad (57)$$

$$J\Omega = J_+\Omega_+ + J_-\Omega_- = Mc^2 \mathit{cos}^2\Phi/2 \quad (58)$$

$$\Rightarrow$$

$$\Omega = c\,\mathit{cos}\Phi/2R_g. \quad (59)$$

Therefore, from eq.(53) & eq.(58), we find that the generalized Smarr formula [2] is also satisfied, that is:

$$Mc^2 = 2(TS+J\Omega). \quad (60)$$

(5) The net scalar angular momentum '$J_+ + J_- = J$' can also be written in terms of the inner and outer event horizon momenta as follows: $J = I\Omega = I_+\Omega_+ + I_-\Omega_- = M_+\Omega_+R_{z+}^2 + M_-\Omega_-R_{z-}^2 = J_+ + J_- = R_g Mc\,\mathit{cos}\Phi$.

## The Zeroth Law

The **Zeroth** Law of Black Hole Thermodynamics, states that the temperature '$T = \hbar c^3 \sin\Phi/8\pi kGM$' is constant over each *stationary* event horizon. Since each *stationary* event horizon has constant $\Phi$, the Zeroth Law is satisfied. This is analogous to the Zeroth Law of Thermodynamics, which states that for any system in thermal equilibrium the temperature is constant throughout the system.

## The First Law [aka Conservation of Energy]

The First Law of Thermodynamics $dE = d(Mc^2) = 2(TdS+\Omega dJ)$ is satisfied, since from eq.(52):

$$dS = 4\pi k N_{\varpi}^2 \cos\Phi d\Phi, \tag{61}$$

from eq.(20):

$$dJ = -R_g Mc \sin\Phi d\Phi, \tag{62}$$

from eqs.(37),(51) & (61):

$$TdS = Mc^2 \sin\Phi \cos\Phi d\Phi/2, \tag{63}$$

& from eq.(59) & eq.(62):

$$\Omega dJ = -Mc^2 \cos\Phi \sin\Phi d\Phi/2. \tag{64}$$

From the above, it is obvious that:

$$2(TdS+\Omega dJ) = 0 \Rightarrow dE = 0. \tag{65}$$

Which corresponds to a black hole of constant mass, i.e.

$$dE = d(Mc^2) = 0. \tag{66}$$

## The Second Law

Since $\Phi \geq 0$, [from eqs. (25) & (26)] then,

$$S = 4\pi k N_{\varpi}^2 \sin\Phi \Rightarrow$$

$$dS = 4\pi k N_{\varpi}^2 \cos\Phi d\Phi \geq 0. \tag{67}$$

The minimum entropy state, where $\Phi = 0$, i.e. $S = 4\pi k N_{\varpi}^2 \sin|0| = 0$, suggests that the parameter '$\Phi$' can be thought of as an analogue of time '$\tau$', where $\mathrm{sgn}(\Phi) = \mathrm{sgn}(\tau)$. So that the entropy always increases when $d\Phi = y d\tau > 0$, and the entropy always decreases in a negative temporal direction when $d\Phi = y d\tau < 0$, where '$y$' is a commensurability mapping function.

## Discussion

Here is a portion of a discussion with a colleague,[6] regarding an article written by Bekenstein [17]. It is included because it serves to illuminate a few additional reasons, other than Third Law consistency, to treat the area of the inner event horizon as a measure of negative entropy.

*"I have followed (from a great distance with great handicap) the cosmologists because I am utterly fascinated by the strides in science they are making with little corroborating experimental verification. I also admire some of what is only informed speculation. However, I do not accept the leap that Bekenstein has made, equivalencing information to entropy.*

*As I understand the argument:*

*Entropy is proportional to the number of arrangements of the smallest building blocks of matter-energy we can count.*

*Information can be coded into a bit pattern.*

*There are conceivably as many bits available as there are the smallest building blocks of matter-energy.*

*Maybe we can store as much information as these small building blocks allow.*

---

(6) A private email communication from Paul E. Schultz, Sent: Thursday, November 06, 2003 7:45 PM

*Therefore information storage capability (not information) is proportional to entropy.*

*Entropy as a quantity is proportional to a black hole's surface area at the no return radius and in short, fulfills all the requirements of the Second Law when things such as radiation take place or when energy goes into the hole. The surface is therefore exactly proportional to the entropy and in fact can be expressed precisely in terms of Planck areas according to Hawking.*

*(So far everything is great and hangs together.)*

*Since information storage capacity is proportional to entropy, ergo, information storage capacity also is on the surface. In fact when something falls into the hole its 'information' somehow gets posted into a binary pattern in the bajillion Planck areas on the surface and becomes the 'hologram' representing everything that fell inside. (Here information storage capacity became the same as information.)*

*This last is to me a great leap that is only supported minimally by intuition.*

*Where's the mechanism??? How does the information flow to the surface of the black hole or any of the other potential holographic surfaces we can imaginatively construct around the universe or any part of it? Why does the counting of these Planck areas mean there are places to 'store' information? (Places, which would have to be ordered to convey information.)*

*(Will we be able to eventually 'read' the surface of a black hole and find out all about everything that fell in or merged with it? What does this tell us about thought processes since they can't be described as a Turing machine process, which I believe all bit-patterned information can.)*

*This idea of physics of the surface being the same physics as the physics we know in the interior depends on being able to be sure that information flow works as described above, right?*

*I hear you, I hear Bekenstein, but I wonder if he didn't make a couple of leaps too far?"*

That's a useful summary of the situation. When examining Kerr black holes, it was found that in order to have black-hole-thermodynamics satisfy the Third Law, the area of inner event horizon has to be proportional to the black hole's negative entropy, which is a measure of its available energy or information content. Even though the area of the inner event horizon is positive, its area is proportional to the hole's negative entropy, which means that the inner event horizon area has to be subtracted from the outer, in order to obtain the total net entropy.

Notice that if the black hole's rotation slows, the inner event horizon's area gets smaller in absolute value, while the outer-event horizon's area gets larger as does the black hole's entropy. If the hole's rotation slows, then the useful energy (negative entropy) gets transformed into entropy. If the hole's rotation slows to a stop, the inner-event horizon's area along with the hole's available energy and information content go to zero, while the outer-event horizon's area approaches its maximum, as does the entropy content of the hole.

A Schwarzschild, or non-rotating black hole, should be maximally entropic, since the area of its outer-event horizon is maximal. It follows that a Schwarzschild black hole should have zero information content, or zero negative-entropy, since the area of the Schwarzschild black hole's inner-event horizon is zero.

It should also be noted, that by using the Penrose mechanism, [18], work can be extracted from rotating black holes, which always have non-zero inner event horizon areas. Whilst no work can be extracted from non-rotating Schwarzschild black holes that have zero inner event horizon areas. Obviously the area of the inner event horizon is related to the black hole's usable energy, which is a function of its negative entropy.

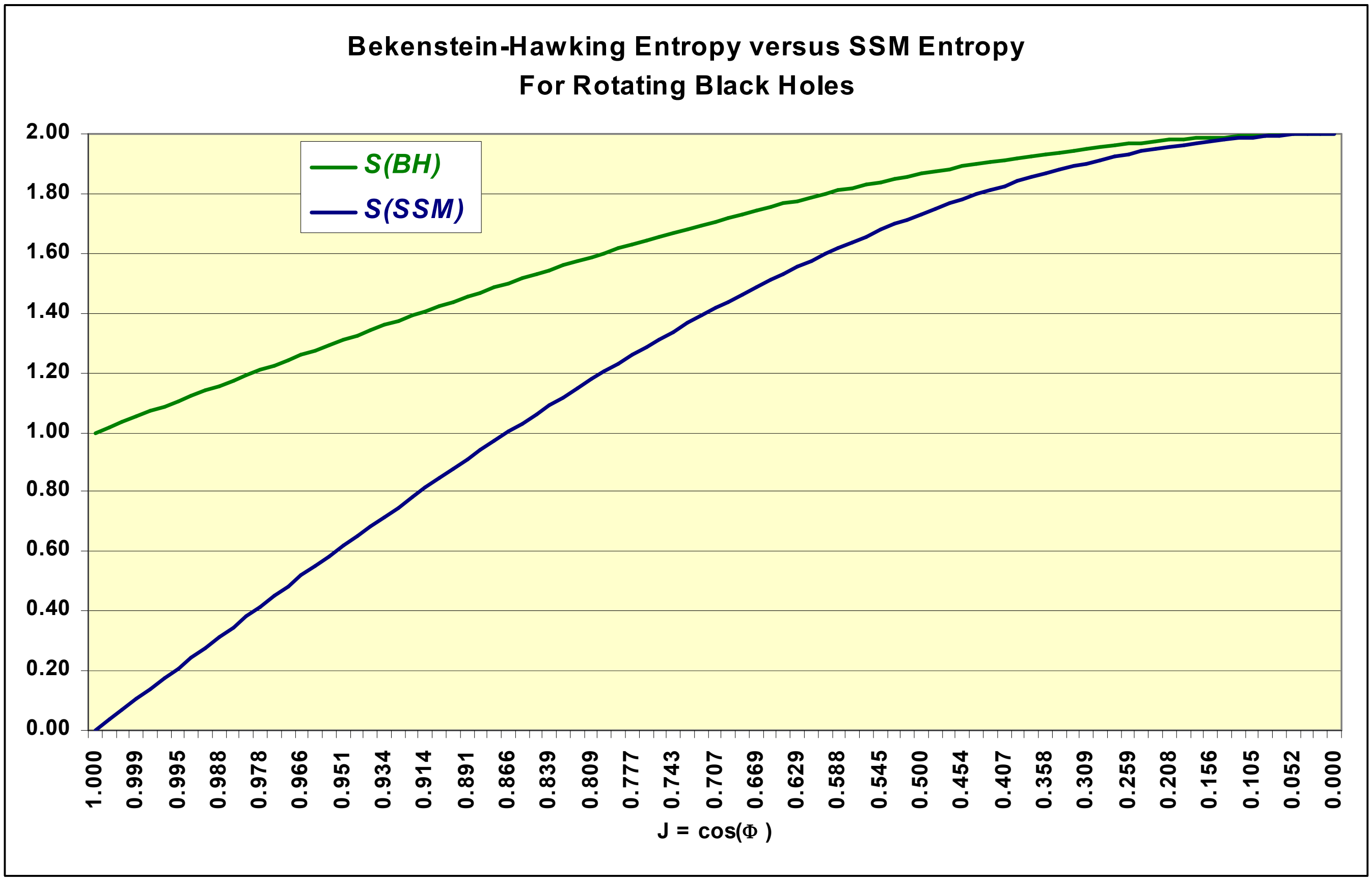

**Figure 1**

Note:

The Bekenstein-Hawking entropy is:

$\mathbf{S_{BH} \equiv S_{+} = kc^3A_{+}/4\hbar G = 2\pi kN_{\varpi}^2\,[1+ \mathit{sin}|\Phi|].}$

The entropy labeled by SSM is:

$\mathbf{S_{SSM} \equiv S = kc^3(A_{+} - A_{-})/4\hbar G = 4\pi kN_{\varpi}^2 \mathit{sin}|\Phi|.}$

The 'SSM' is in reference to the treatment of Kerr black hole entropy in the 'Super Spin Model' which is discussed in [11], [12] & [13].

The Kerr black hole's angular momentum amplitude is:

$\mathbf{J = I\Omega = I_{+}\Omega_{+} + I_{-}\Omega_{-} = M_{+}\Omega_{+}R_{z+}^2 + M_{-}\Omega_{-}R_{z-}^2 = J_{+} + J_{-} = R_gMc\mathit{cos}(\Phi) = \hbar N_{\varpi}^2\,\mathit{cos}\Phi.}$

The above graph is scaled by:

$S_{BH} \equiv S_{BH}/2\pi kN_{\varpi}^2$

$S_{SSM} \equiv S_{SSM}/2\pi kN_{\varpi}^2$

$J \equiv J/\hbar N_{\varpi}^2$

**Acknowledgements**

I am very grateful for useful discussions with Paul Schultz, and for the proof reading and editing by my wife Connie. Above all, I wish to acknowledge and thank the eternal *"God (Who) used beautiful mathematics in Creating the World!"* as P.A.M. Dirac proclaimed.